\documentclass{pnastwo}

\url{www.pnas.org/cgi/doi/10.1073/pnas.0709640104}
\copyrightyear{2014}
\issuedate{Issue Date}
\volume{Volume}
\issuenumber{Issue Number}

\newcounter{note}

\begin{document}

\title{Understanding the spreading power of all nodes in a network: a continuous-time perspective}

\author{Glenn Lawyer\affil{1}{Max Planck Institute for Informatics, Saarbr{\"u}cken, Germany}}

\contributor{Submitted to Proceedings of the National Academy of Sciences
of the United States of America}

\significancetext{ 
Memes, innovations, diseases, invasive species-- many socially relevant phenonenon are types of spreading processes. Despite the ubiquity and social impact of such processes, and the common use of network models to understand their dynamics, we still do not have a theoretically sound approach for quantifing how structural features of the source node influence the outcome of such processes. By taking a fundamentally new approach to defining node influence, the current work teaches us that influence is a balance between self and neighbor degree. Nodes with low influence gain what strength they have from their neighbors onward connectivity, with the balance shifting to self degree as influence increases. This shift is accentuated by the higher triangle count in denser networks.
}

\maketitle

\begin{article}
\begin{abstract}{
Centrality  measures such as the degree, k-shell, or eigenvalue centrality can identify a network's most influential nodes, but are rarely usefully accurate in quantifying the spreading power of the vast majority of nodes which are not highly influential. The spreading power of all network nodes is better explained by considering, from a continuous-time epidemiological perspective, the distribution of the force of infection each node generates. The resulting metric, the \textit{Expected Force} (ExF), accurately quantifies node spreading power under all primary epidemiological models across a wide range of archetypical human contact networks. When node power is low, influence is a function of neighbor degree. As power increases, a node's own degree becomes more important. The strength of this relationship is modulated by network structure, being more pronounced in narrow, dense networks typical of social networking and weakening in broader, looser association networks such as Internet webpages. The ExF can be computed independently for individual nodes, making it applicable for networks whose adjacency matrix is dynamic, not well specified, or overwhelmingly large.
}\end{abstract}


\abbreviations{FoI, force of infection; ExF, expected force}

\dropcap{N}etworks have become the premier approach to describing spreading processes 
such as epidemics or information diffusion because they express the heterogeneity of interactions characteristic of many human activities~\cite{danon2011}. 
Thirty years of innovation have refined our ability to identify nodes which are highly influential to the outcome of almost any spreading process on a given network  via features such as betweenness centrality~\cite{Freeman1979,Friedkin1991}, eigenvalue centrality~\cite{Bonacich1987}, degree~\cite{Albert2002}, or k-shell~\cite{kitsak2010}.
Yet highly influential nodes are rare by definition,  
and the just listed measures are not 
informative for the vast majority of network nodes.
 These centrality measures only rank nodes and are not designed to quantify spreading power~\cite{kitsak2010,Bauer2012,Estrada2005}. 
While the rankings accurately identify the few highly influential nodes, they 
can considerably underestimate the spreading power of non-hub nodes~\cite{Sikic2013}.  
 Nor do these rankings incorporate the dynamics of spreading processes~\cite{Borgatti2005,Klemm2012}.
This leaves open the question of quantifying the spreading power of the vastly more numerous non-highly influential nodes, and indeed understanding the nature of node spreading power itself.
As highly influential nodes only rarely originate spreading processes, be they pathogenic disease~\cite{Taylor2001,Reperant2010}, innovative ideas~\cite{Christensen1997}, or chatter~\cite{Meeyoung2010}, 
there is deep intellectual hunger and practical utility in accurately measuring and understanding the spreading power of each individual node in a network.

A node's spreading power is the force with which it can push a spreading process to the rest of the network. 
More precisely, in a susceptible-infected (SI) spreading process without recovery, which inevitably reaches the entire connected component of the network, the spreading power of the seed node predicts the delay before half (or some other large percentage of) the network is reached.
In a process with recovery to either the susceptible (SIS) or immune (SIR) state, spreading power correlates to the probability that a node can seed an epidemic
given that the ratio of the  per-contact transmission rate to the rate of recovery allows for, but does not guarantee, an epidemic.
When this ratio exceeds this critical range, the dynamics approach the SI system as a limiting case.

Several approaches to quantifying the  spreading power of all nodes have recently been proposed,
including the \textit{accessibility}~\cite{Travencolo2008,Viana2012}, 
the \textit{dynamic influence}~\cite{Klemm2012},
and the \textit{impact}~\cite{Bauer2012} (See supplementary \stepcounter{note} Note S\arabic{note}).
 These extend earlier approaches to measuring centrality by explicitly incorporating spreading dynamics,
and have been shown to be both distinct from previous centrality measures and  more highly correlated with epidemic outcomes~\cite{Bauer2012,Klemm2012,daSilva2012}.
Yet they retain the common foundation of the more usual approaches to centrality, counting walks on the network \cite{Borgatti2005,Borgatti2006,Estrada2010, Benzi2013}.
As the walks are counted using powers of the adjacency matrix, spread is observed only in discrete time.

Epidemiology, in contrast, studies the continuous-time dynamics of 
 the force of infection (FoI), defined as the current rate at which susceptible nodes are becoming infected~\cite{anderson1992}. 
In network models, the FoI is directly proportional to the current number of edges between infected and susceptible nodes. 
The critical distinction between FoI and walks is that the FoI is determined by the number of infected-susceptible edges, independent of their distance from the seed node.
The critical distinction between continuous- and discrete-time is that  continuous-time
allows resolution down to the  first few transmissions, a level not easily expressed in a discrete-time framework where multiple transmissions may occur at each time step
\stepcounter{note} (Note S\arabic{note}).
The distinction is acute, as the number of events per time step grows at a double-exponential rate in scale-free networks~\cite{fountoulakis2012}, the type of network most representative of human social structures~\cite{Barabasi1999} and perhaps even life itself~\cite{Almaas2004}.

The continuous-time epidemiological perspective suggests that node spreading power can be accurately quantified by appropriately summarising the distribution of the number of infected-susceptible edges after a small number of transmission events arising from a seed node in an otherwise fully susceptible network; that is, by the expected FoI generated by that node.
We here propose such a measure, named the \emph{Expected Force} (ExF), and show that it  outperforms the accessibility, k-shell, and eigenvalue centrality 
\stepcounter{note}(Note S\arabic{note})
\stepcounter{note}
in predicting epidemic outcomes in SI, SIS, and SIR spreading processes,
in both discrete- and continuous-time.
The ExF's basis in local neighborhood structure means that it is applicable even when the full adjacency matrix is either unknown or inherently unknowable.
The metric naturally  extends to weighted and directed networks.
Most importantly, the ExF is able to illuminate the factors responsible for node spreading power.


\section{Definition and validation}


The ExF is a node property derived from local network topology, independent of the rest of the network or any specific spreading process. It is formally defined as follows. Consider a network with one infected node $i$ and all remaining nodes susceptible.
Enumerate all possible clusters  $J=1, \ldots, |J|$ of infected nodes after $x$ transmission events, assuming no recovery (See Fig 1).
Generally speaking, $x=2$ is sufficient and assumed for the rest of this manuscript.
Hence $J$ includes all possible combinations of $i$ plus two nodes at distance one from $i$, and $i$ plus one node at distance one and one at distance two.
The enumeration is over all possible orderings of the transmission events.
 Two neighbors of the seed ($a$ and $b$) form  two clusters 
($[i \rightarrow a, i \rightarrow b$] and $[i \rightarrow b, i \rightarrow a$])
or, if $a$ and $b$ also share an edge, four clusters.
Define the degree of a cluster of nodes $d_j$ as the number of edges connecting nodes  inside the cluster to nodes outside. 
After two transmissions without recovery, the FoI of a spreading process seeded from node $i$ is a discrete random variable taking a value in $(d_1, \ldots, d_J)$, allowing for the proportionality constant equal to the transmission rate of the process.
The \emph{expected} FoI can be approximated by the entropy of the $d_j$ after normalisation
\begin{equation}
ExF(i)=- \sum_{j=1}^{J} \bar{d_j} \log(\bar{d_j})
\end{equation}
 where $i$ refers to the seed node and  $\bar{d_k}=\frac{d_k}{\Sigma_Jd_j},\, \forall k \in J$.
 The entropy is needed for generating the expected value due to the extreme variability in the shape, number of modes, and number of terms in the distributions of $d_j$ for different seed nodes.

Setting $x=2$ is recommended but not required.
Increasing the number of transmissions beyond two adds very little information while increasing the computational cost 
\stepcounter{note}(Note S\arabic{note}), in agreement with  other proposed spreading power metrics~\cite{Bauer2012,Klemm2012} and 
consistent with the decaying influence of longer paths in the calculations of the eigenvalue, subgraph, and related centralities \cite{Bonacich1987,Estrada2005,Estrada2010, Benzi2013}.
In certain cases, however, it may be desirable to consider more transmission events. For example, a node at the end of a chain of length two can only form one transmission cluster of size two, hence its ExF is zero. 
Comparing two such nodes requires setting $x=3$, in which case a subscript can be used for clarity (e.g. ExF$_3$).

One modification may be in order for SIS/SIR processes, inspired by the following. Imagine a node with degree one connected to a hub. While such a node will have a high ExF, its chance of realizing this force depends entirely on transmitting to the hub before recovery. 
Such nodes are common in dense social networks. For example, 84\% of the 225K nodes in an EU institution email network~\cite{Leskovec2007} have degree one.
In such networks,
it may be helpful to 
multiply the ExF by the log of the seed node's degree
 after first rescaling the seed's degree by some factor $\alpha>1$.
\begin{equation}
ExF^{M}(i)= \log(\alpha\, deg(i))\, ExF(i)
\end{equation}
The rescaling is motivated in that the log of one is zero, and the ExF$^M$  is most informative in networks where many nodes have degree one.
The rescaling factor must be greater than one, and should also be small to avoid overpowering the influence of the degree.
 In the rest of this manuscript, we use $\alpha=2$, the smallest integer which satisfies these criteria. 
Computing the ExF$^M$ for $\alpha$ ranging from 1.0001 to 16 does not substantively alter the metric, as all such variations show correlations greater than 0.99 to ExF$^M$ computed with $\alpha=2$ \stepcounter{note}(Note S\arabic{note}).


Running times are discussed in \stepcounter{note} Note S\arabic{note}.
Example code providing an implementation of the ExF is available at\\
https://github.com/glennlawyer/ExpectedForce.

\begin{figure}[t]
\centerline{\includegraphics[width=.45\textwidth]{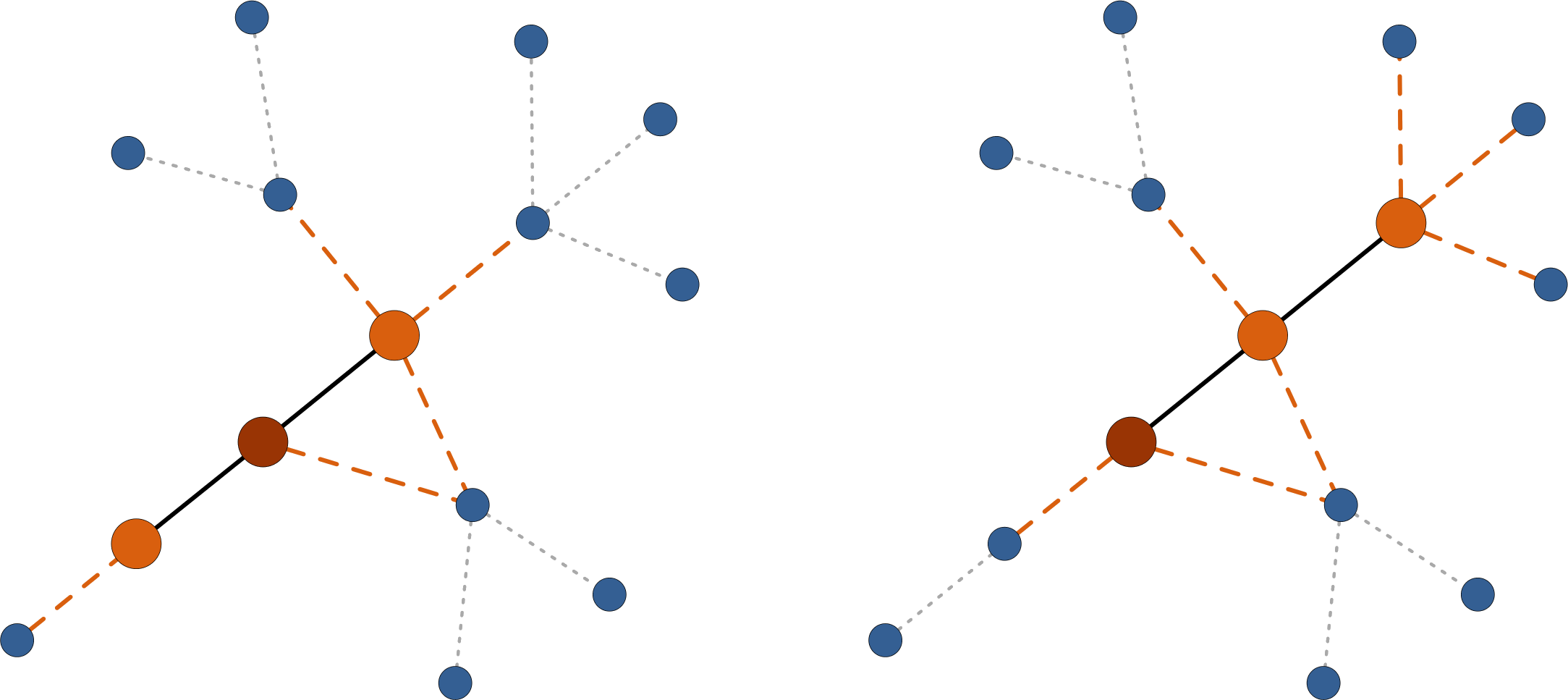}}
\caption{This network will be in one of eight possible states after two transmissions from the the seed node (red). Two states are illustrated, where the seed has transmitted to the two orange nodes along the solid black edges. Each state has an associated number of (dashed orange) edges to susceptible nodes (blue), the cluster degree. States containing two neighbors of the seed can form in two ways or, if they are part of a triangle, four ways. The eight network states associated with the pictured seed node arrise from thirteen possible transmission clusters. The ExF of the seed node is the entropy of the distribution of the normalized cluster degree over all possible transmission clusters.}
\end{figure}

\subsection{Correlation to epidemic outcomes}
We measure correlations between ExF and epidemic outcomes 
on five families of simulated networks chosen such that their densities and degree distributions span a wide range of human contact structures (Methods, and Table S4).
One hundred random networks of 1,000 nodes are generated in each family. Further comparison is made using a suite of twenty four empirical networks ranging from 1,133 to 855,800 nodes (Table S5).
Epidemic outcomes
are observed by simulating multiple epidemics in both continuous and discrete time from a number of seed nodes in each network.
Correlations are measured between these outcomes and the ExF, ExF$^M$,
accessibility, eigenvalue centrality, and the k-shell of the seed nodes 
(Note S3).

\begin{figure*}[t]
\begin{center}
\centerline{\includegraphics[width=.8\textwidth]{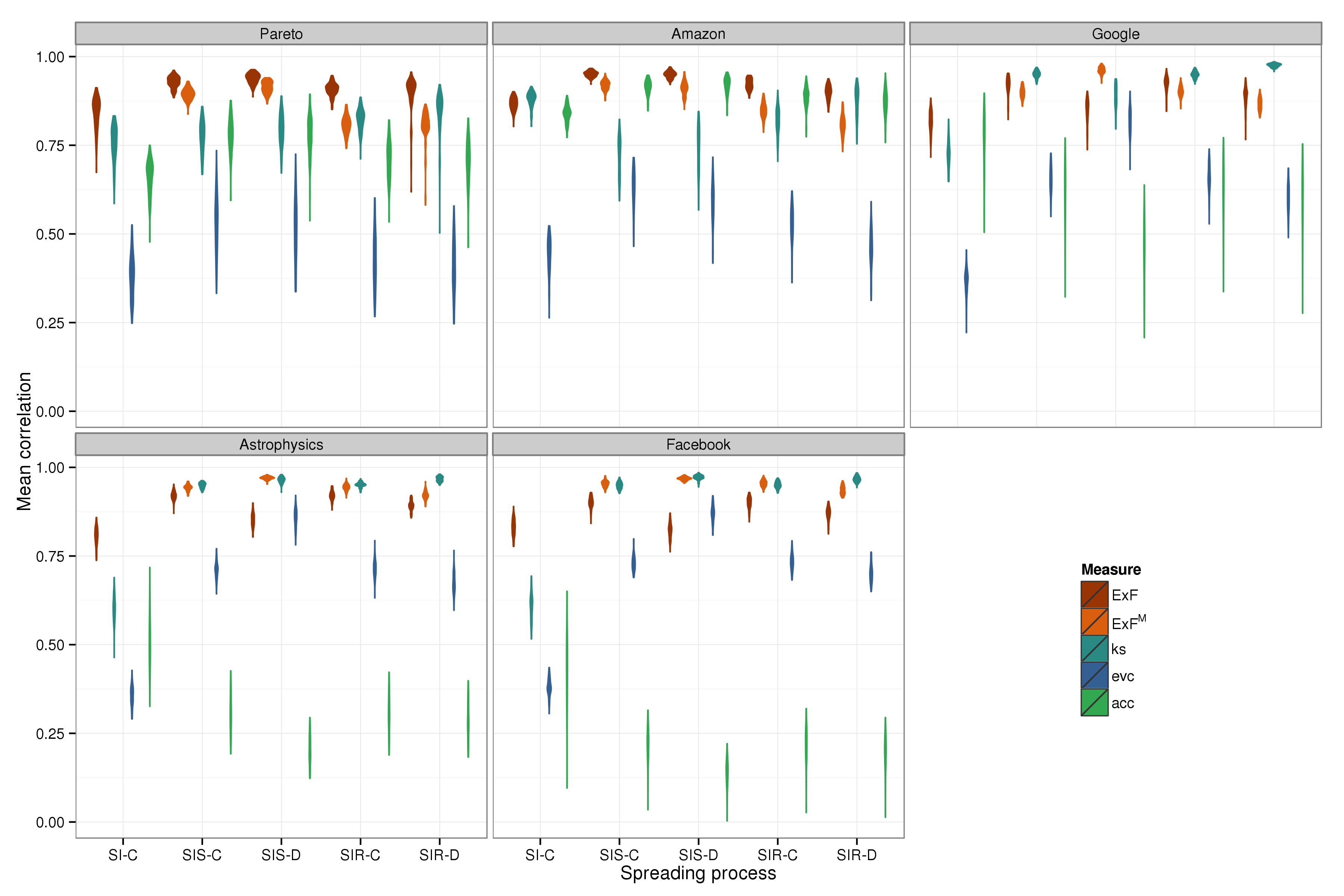}}
\caption{Observed correlations between metrics and epidemic outcomes in each network family. The ExF and ExF$^M$ (orange shades) are consistently strong, with mean correlations greater than 0.85 and small variance. The other measures (k-shell, eigenvalue centrality, and accessibility, blue-green shades) show both lower mean values and higher variance. Each violin summarizes correlations computed on 100 simulated networks. Spreading processes (x axis) are suffixed to indicate simulations in continuous (-C) or discrete (-D) time. }
\end{center}
\end{figure*}

The ExF is highly predictive of all epidemic outcomes on all networks tested, simulated and empirical. Mean correlation to SI process outcomes is 83\% on simulated and 74\% on empirical networks. For processes with recovery, mean correlation is 91\% on simulated and 82\% on empirical networks. Standard deviations  over the one hundred simulated networks in each family are typically $0.02-0.03$. The 95\% confidence bounds on empirical networks are in the same range.
In almost all cases the ExF significantly outperforms the  accessibility, eigenvalue centrality, and the k-shell (difference in mean correlations greater than the standard deviation of the higher mean).
The ExF is only bested in the following three cases, each time by the k-shell: discrete time SIR processes in the simulated Astrophysics (0.97 vs 0.92) and Facebook (0.97 vs 0.94) networks and the SI process on the ``email-EUAll'' network (0.50 vs 0.41).
The performance of the k-shell was surprisingly strong, given that two previous studies by independent groups have observed rather poor performance for this metric~\cite{Klemm2012,daSilva2012}.
The observed correlations on 100 simulated networks in each family are shown in  Fig. 2 (also Table S6). 
The measured correlations and their standard errors for all empirical networks are shown in Fig. 3 (also Tables S7-9).

 The ExF's predictive power is robust to variation in network structure.
The theory behind the ExF$^M$ suggests that the ExF might perform less well for SIS/SIR processes on denser networks, yet 
mean correlation for continuous time SIS processes is barely changed between the loose Pareto/Amazon networks (0.93/0.95) and the dense Astrophysics/Facebook networks (0.92/0.90).
As expected, the predictive power of the ExF$^M$ improves on the denser networks 
(mean correlations: Pareto/Amazon 0.89/0.92, Astrophysics/Facebook 0.94/0.95).
The accuracy of the accessibility metric, in contrast, collapses for all spreading processes on the dense networks (mean correlation over all spreading processes:  Pareto/Amazon 0.74/0.90, Astrophysics/Facebook 0.28/0.20.)
A previous analysis  which observed similar poor performance for the accessibility on dense networks  concluded that spreading processes seeded from  nodes with low accessibility are not capable of entering the epidemic phase~\cite{daSilva2012}. 
Our results show this is not the case, as these nodes have a small yet observable epidemic potential which the ExF is able to capture and quantify.
Performance of the k-shell and the eigenvalue centrality is also strongly influenced by network structure. For SIS/SIR processes, both showed higher mean and sharply reduced variance on the denser networks. 
In an interesting contrast, the k-shell's predictive power for SI processes is reduced in denser networks.
The eigenvalue centrality's performance also varies by spreading process, showing its best performance on discrete time SIS models-- though again this variation is modulated by network density.
Two other independent groups have observed that relationships between centrality rankings and epidemic outcomes are strongly influenced by network structure and  the parameters of the spreading processes~\cite{Sikic2013,Bauer2012},
leading the authors of \cite{Sikic2013} to conclude that these measures severely underestimate the epidemic impact of structurally peripheral nodes.

\subsection{Weighted graphs}
\hspace{1ex} The ExF generalizes to graphs with weighted edges, where we assume the edge weights correspond to per-edge transmission likelihoods. Use these  weights to calculate the probability of each way that each cluster could occur, and re-define the cluster degree as the sum of all edge weights leading out from that cluster. The extension to directed graphs is also straightforward; limit the 
enumeration to edges leading from an infected to a susceptible node. 

We test this generalization by computing the weighted and unweighted ExF on 1,000 node networks with Pareto (1,2.3) degree distributions and edge weights chosen according to one of the following 
three distributions:
uniformly distributed between one and three, 
uniformly distributed between one and ten,
and exponentially distributed with unit rate, weights rounded up to the nearest integer.
Fifty networks were simulated for each distribution of edge weights.
Correlation between the weighted and unweighted ExF was greater than 0.99 for all network edge weighting distributions tested.
As expected from the tight correlation, the weighted and unweighted ExF  showed no meaningful difference in predictive ability, which remained high.
Observed correlations between node ExF and epidemic potential in discrete-time SIS processes were 0.88/0.89 $\pm 0.03$ (unweighted/weighted ExF) under the uniform-3 scheme,
0.83/0.04 $\pm 0.03$ under the uniform-10 scheme, and
0.80/0.79 $\pm 0.05$ under the exponentially distributed weighting scheme.



\begin{figure*}[t]
\begin{center}
\centerline{\includegraphics[width=.8\textwidth]{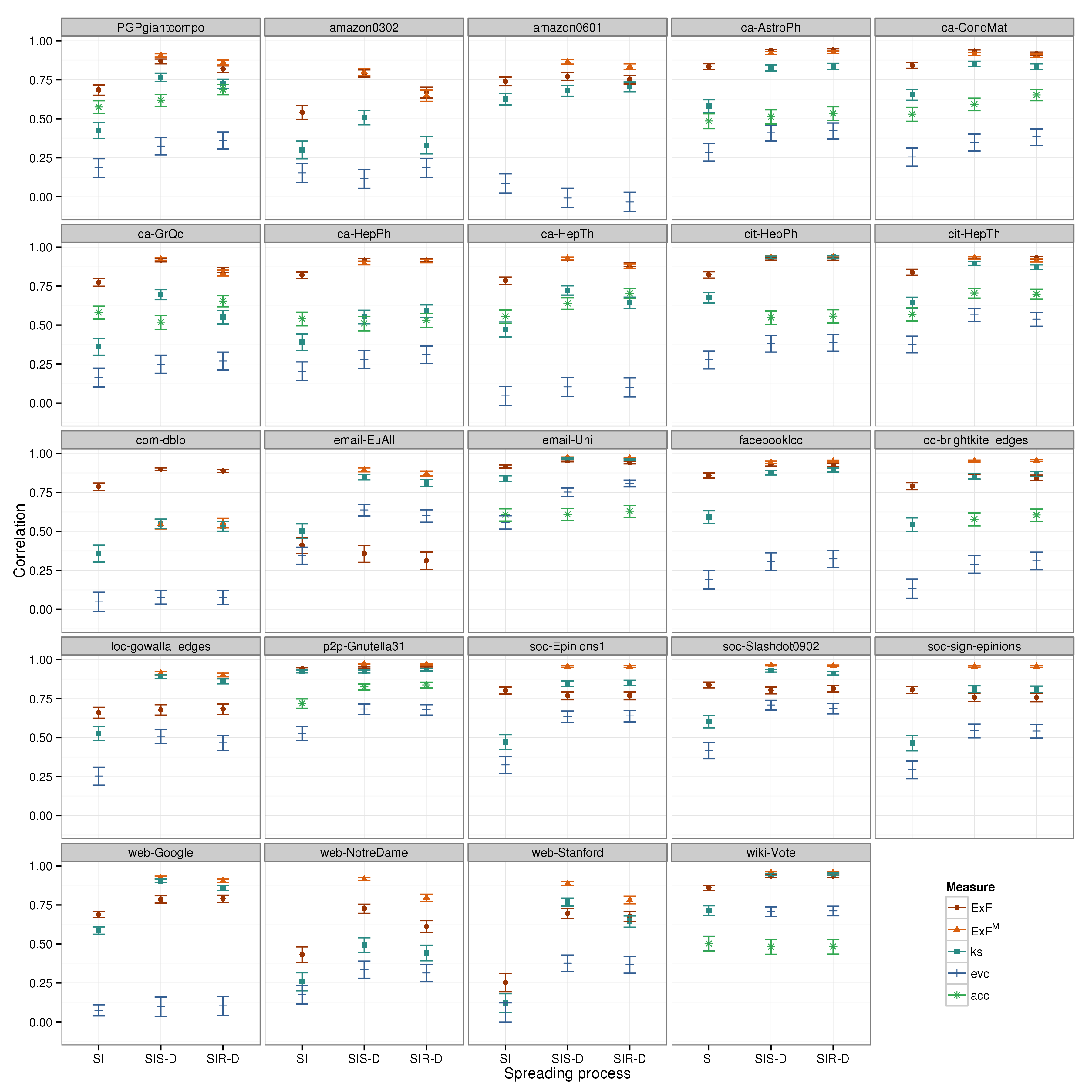}}
\caption{Observed correlations and 95\% confidence intervals between metrics and epidemic outcomes on the 24 empirical networks. The ExF and ExF$^M$ (orange shades) show strong performance, consistently outperforming the other metrics (k-shell, eigenvalue centrality, and accessibility when computed, blue-green shades).  The suffix ``-D'' indicates spreading processes simulated in discrete time. }
\end{center}
\end{figure*}

\section{Discussion}

The ExF predicts all types of epidemic outcomes with high accuracy over a broad range of network structures and spreading processes. The low variance in observed correlations over multiple simulated network and epidemic models shows that the measure is robust, as do the tight confidence bounds on empirical  networks.
What, then, does it tell us about the nature of node spreading power?
The definition of the ExF implies that spreading power is determined by both the degree of the node and the degree of its neighbors, and that the relative influence of these two factors is different for nodes of low versus high spreading power. 
Weaker nodes gain what strength they have from their neighbors onward connectivity, whereas more influential nodes get their strength from their large number of connections. 
These relationships are accentuated by network density.

This is a result of the combinatorics behind the enumeration over transmission clusters.
Clusters are formed either by choosing two neighbors from the first geodisc, a term which grows approximately quadratricly in node degree, or by each path leading to the second geodisc, a term which grows approximately linearly in the size of this disk. 
Generally speaking, the second geodisc is much larger than the first.
Hence spreading power is determined largely by terms from the second geodisc until the quadratic growth of the influence of the first overtakes it.
The influence of network density reflects the ExF's sensitivity to 
network motifs such as triangles and squares.
Each triangle touching the seed node contributes four terms to the enumeration $J$, 
meaning the contribution of the first geodisc grows faster than quadratically.
Sensitivity to squares and triangles is also granted by the entropy term.
These motifs, which are more common towards the cores of dense networks, reduce the disparity of the cluster degree distributions thus increasing the ExF
(Fig. 4). 

\begin{figure}[t]
\begin{center}
\centerline{\includegraphics[width=.4\textwidth]{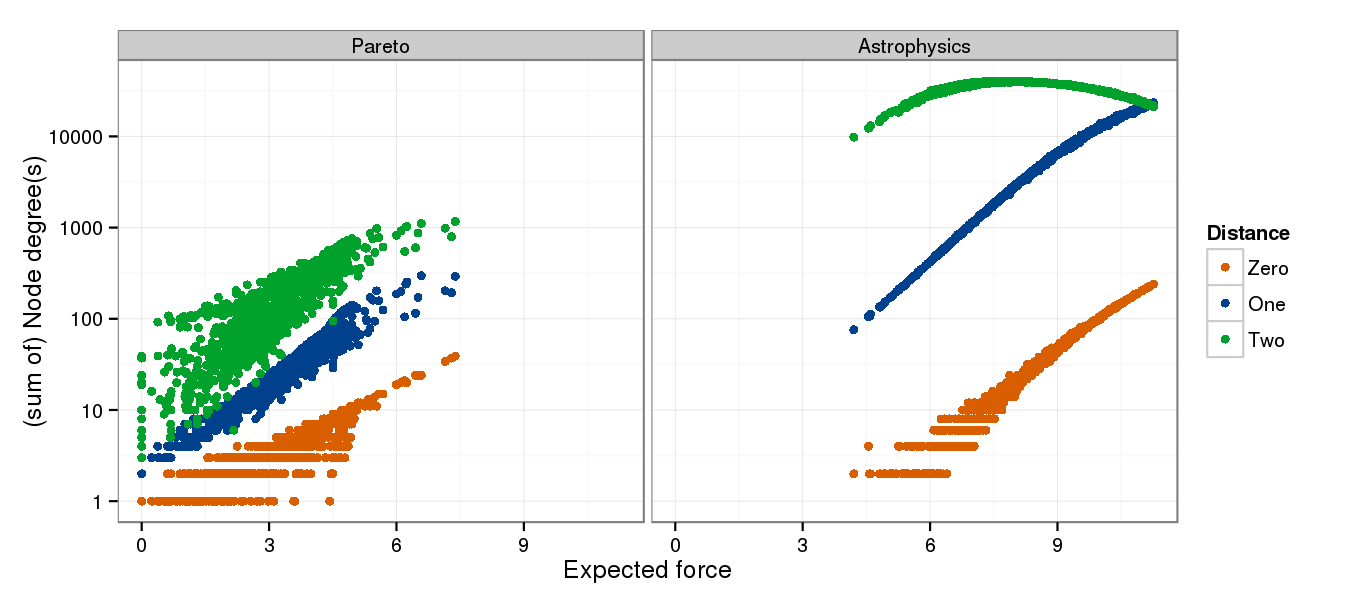}}
\caption{Plotting ExF (x-axis) versus node degree (orange), the sum of the degree of all neighbors (blue), and the sum of the degree of all neighbors at distance 2 (green)  shows that for nodes with low ExF, the neighbor's degree has strong correlation to ExF, while for nodes with high ExF their own degree is more closely correlated  ways. The result is accentuated in denser collaboration networks in comparison to more diffuse Pareto networks.}
\end{center}
\end{figure}

The approach taken by the ExF is fundamentally different than that taken by most centrality measures.
Centrality measures typically set out to produce a ranking which identifies the most influential nodes in the network, under the assumption that highly influential nodes are those with the maximal sum of some type of walk~\cite{Bauer2012,Borgatti2005,Borgatti2006,Estrada2010,Benzi2013}.
The choice of the appropriate type, scaling, and length of walks 
contain implicit assumptions regarding network flows \cite{Borgatti2005}, cohesion structure~\cite{Borgatti2006}, and/or other topological characteristics of the network~\cite{Estrada2010,Benzi2013}.
The k-shell is a slight exception, as it was originally intended to precipitate out the most cohesive regions of the network rather than to explicitly rank nodes within cohesive regions~\cite{Seidman1983}, yet it is now recognized as one of the best centrality measures for identifying a network's most influential spreaders~\cite{kitsak2010}.
Spreading power metrics generalize the walk counting framework by explicitly including transmission probabilities when scaling the walks \cite{Bauer2012,Klemm2012,Travencolo2008,Viana2012}.
The question not asked is if the type, scaling, and lengths of walks best suited to identifying the most important nodes applies equally well to the rest of the network.
To the extent that the optimal choice of factors depends on network topology, then the difference in topology between core and periphery suggests that choices well suited to the core are seldom appropriate for the remainder of the network.

 Both the combinatorics behind the ExF and the walk counting behind most centrality measures agree that 
influential nodes are those which combine high degree with a preponderance of influential neighbors.
The ExF has high rank correlation with both the eigenvalue centrality and the k-shell (0.62-0.92 across the simulated network families \stepcounter{note} (Note S\arabic{note} and Table S2).
 Likewise, the ExF has 60-90\% agreement with the eigenvalue centrality on the top ten network nodes  and 100\% agreement with the k-shell.
The difference between walk counting and the ExF is that the ExF adapts the relative influence of different walks and walk lengths based on local connectivity, whereas approaches based on functions of the adjacency matrix apply a fixed protocol.
The eigenvalue centrality is weighted node degree, where the weights are the importance of the neighbors \cite{Bonacich1987,Estrada2005}. 
But the eigenvalue centrality is strictly a global measure, unable to distinguish more subtle variations in local structure \cite{Estrada2005,Benzi2013}.
The k-shell erodes node degree to match the number of neighbors with similar degree. Since this discards remaining information on the individual degree of nodes within a common shell, the accuracy of its predictions is heavily influenced by the number of shells in the network.
The accessibility combines node and neighbor degree into a measure of the number of nodes likely to be reached by walks of a given length~\cite{Viana2012}.
But this approach has difficulties quantifying nodes in dense, small diameter networks, which accentuate differences between core and peripheral topology.

Calculation of the ExF depends only on the local topology. 
This allows epidemic outcomes on the whole network to be predicted with high accuracy even when only a small portion of the network is known \stepcounter{note}(Note S\arabic{note}).
It is rare for the full structure of a real network to be fully known; typically the network structure is inferred from indirect, incomplete, and often biased observations.
Specification of an adjacency matrix is even more difficult when the underlying network is dynamic.
In contrast,
estimates of eigenvalue centrality fluctuate depending on which nodes are sampled~\cite{Costenbader2003}. Both the  pagerank~\cite{Ghoshal2011} and the k-shell~\cite{Adiga2013} are highly sensitive to pertubations in network topology, making them unreliable for incomplete or noisy systems.
Reliance on a local network does, however, lead to one weakness in the ExF.
A network may contain large but disparate communities. 
Here, a node serving as a bridge between two communities 
might be able to spread a process to the entire network with more force than a node far from the bridge, even when the second node has more (local) spreading power than the bridging node.
The ExF's local nature makes it blind to these larger topological constraints on spread.




The ExF is strongly correlated to epidemic outcome, outperforming existing metrics of node spreading power and centrality. 
The measure depends only on local network topology, allowing its use in dynamic as well as static networks.
For most nodes, the most important determinant of their spreading power is the sum of their neighbors' degree. As node power grows, so does the importance of the node's own degree. This relationship is accentuated in denser networks.

\section{Methods}

\subsection{Epidemic simulations and outcomes}
The epidemic outcome for SI process is the time to half coverage (\textit{tthc}); the time until exactly half the nodes in the network are infected. 
This is measured at each seed node \stepcounter{note}(Note S\arabic{note}) by simulating 100 spreading processes and fitting the observed \emph{tthc} from all simulations to a gamma distribution.
Simulations are run in continuous time, with the time of the next transmission event drawn from an exponential distribution with rate equal to the current number of infected-susceptible edges. 
Discrete time does not provide sufficient resolution for SI processes.

Spreading processes with recovery are of interest when  the ratio of transmissibility to recovery for the process  $\beta$ is in the critical range which allows for but does not guarantee an epidemic.
The epidemic outcome is a node's epidemic potential (\textit{EPo}), the probability that a given node can seed an epidemic \stepcounter{note}(Note S\arabic{note}). 
The \textit{EPo} is measured by simulating 100 outbreaks per seed node and counting the number which result in at least half of the network nodes becoming infected. 
 Continous-time simulations model the time of the next transmission as in the SI model just described, 
except the transmission rate parameter scales the current number of infected-susceptile edges by some $\beta$ in the critical range (see below), 
and  the time of the next recovery from a unit rate exponential distribution weighted by the number of infected individuals.
In each round of the discrete-time simulations,  transmission occurs along each infected-susceptible edge with probability $r=-\log(1-\beta)$ (standard conversion of rate to probability) and nodes recovering at the end of the round.

The critical range for $\beta$ can be defined empirically using the criterion that if $\beta$ is in the critical range, then a large majority of nodes will have \textit{EPo} $\in [2\%, 98\%]$.
We here set $\beta$ independently for each network to a fixed multiple of $1/\lambda$, where $\lambda$ is the largest eigenvalue of the adjacency matrix.
Similarity in network structure allows the same multiple to be used for both the Pareto and Amazon networks, likewise the Astrophysics and Facebook networks. 
Internet hyperlinks lie between these two other classes, and the multiples from the social networks yields good results.
This proceedure resulted in at least 96\% of nodes in continuous-time simulations with \emph{EPo} in the critical range; in discrete time the relevant figure is 76\% or better (Table S3).

\subsection{The networks}
The five families of simulated networks are defined by their degree distributions, 
one theoretical (Pareto), and four derived from the following empirical human contact networks:
the Amazon co-purchase network from May 2003 ~\cite{Leskovec2007a}, 
the graph of Internet hyperlinks from the 2002 Google programming contest~\cite{Leskovec2009},
the collaboration network from ArXiv Astrophysics between 1993 and 2003~\cite{Leskovec2007}, and
Facebook wall posts from the New Orleans network~\cite{viswanath2009} (Table S4).
The Pareto and Amazon networks are characterized by large diameter and low density. 
The Astrophysics and Facebook networks are two orders of magnitude more dense and have correspondingly smaller diameter.
Google's map of Internet hyperlinks lies in between the other two families.
The networks can also be characterized by the largest eigenvalue of their adjacency matrix, as theory suggests that the critical disease transmission probability dividing epidemic from extinction regimes is the inverse of this value~\cite{Klemm2012}, which further implies that a network's inherent susceptibility to disease spreading is reflected by this eigenvalue.
Again, the selected network families cover a wide range of inherent susceptibility to disease.
Simulations are conducted using giant components of 1,000 nodes.
Networks with a Pareto (1,2.3) degree distribution are simulated using the Chung Lu protocol, a versatile and general method for constructing random graphs with given expected degree sequence~\cite{Chung2002}.
The remaining networks are simulated by randomly sampling 1,000 values from the degree sequence of the actual graph without replacement and generating a graph from these values using the Igraph function ``degree.sequence.game'' which generates undirected, connected simple graphs matching the input degree sequence \cite{igraph}.


Empirical networks were selected and downloaded from the Stanford Large Network Repository (SNAP) and  Alex Arenas's  collection according to the following criteria:  
having between 1,000 and 1,000,000 nodes in the largest connencted component, 
representing one clear network structure, 
and, in the case that the same network is sampled at multiple timepoints, the latest timepoint. 
Twenty one networks from SNAP and two from Alex Arena passed these criteria.
The simulated Amazon networks are derived from the earliest Amazon co-purchase network in SNAP.
For completeness, this network is also included in the suite of empirical networks.
The  Facebook network  was downloaded from the Max Planck Institute for Software Systems.
For the purpose of testing, networks are treated as undirected graphs with multiple and self-edges removed.
The 24 networks are characterized in Table S5, which includes the internet address of the collections.

The size of the empirical networks required slight modifications to the overall approach.
Seed nodes are 1,000 nodes selected uniformly at random.
Epidemics with recovery are simulated only in discrete time. 
As can be seen from the results on the random networks, discrete time simulations provide approximately the same mean outcome as continuous time, and only slightly higher variance. 
The transmission/recovery probability ratio $\beta$ is determined independently for each network  by binary search over possible values until one is found such that a minimum of 80\% of tested nodes have \emph{EPo} between 0.05 and 0.95.
When the network has more than 25,000 nodes, the \textit{tthc} is measured as the time of the 1,000$^{th}$ transmission rather than the time when half the network is infected. 
Finally, the R software package does not gracefully handle the multiplication of matrices larger than 25Kx25K, even with the Matrix package~\cite{Bates2012}. 
Hence the accessibility was not computed for networks with more than 25k nodes.

\begin{acknowledgments}
This work was supported out of the general budget of the Max Planck Society.
\end{acknowledgments}

\bibliographystyle{pnas}
\bibliography{cleanbib}

\end{article}

\end{document}


\title{Supplementary materials for ``Understanding the spreading power of all nodes in a network: a continuous-time perspective.''}

\author{Glenn Lawyer}
\affiliation{Max Planck Institute for Informatics} 

\maketitle
\clearpage
\setcounter{page}{1}
\newcounter{note}

\noindent
\textbf{The Notes and Tables S1-3 support claims made in the main text. Tables S4-5 describe the networks used in the study. Tables S6-9 give the results of the evaluations in numerical form, duplicating the figures in the main text.}

\vspace{2em}
\stepcounter{note} \noindent
\textbf{Note S\arabic{note}. Incorporating the dynamics of spread.}
Borgatti's 2005 classification of centralities in terms of network flows 
observes that centralities often give the ``wrong'' answer when applied to a flow which they do not model, and further claims that many sociologically interesting processes are not well described by existing centralities~\cite{Borgatti2005}. Spreading processes ranked high on this list.
Since then, several approaches have been published which explicitly seek to incorporate the dynamics of spread into the walk count.
These include the \textit{accessibility}~\cite{Travencolo2008,Viana2012}, 
the \textit{dynamic influence}~\cite{Klemm2012},
and the \textit{impact}~\cite{Bauer2012}.
The accessibility is a modified form of hierarchical degree which controls for both transmission probabilities and the diversity of walks of a given fixed length~\cite{Viana2012}.
The dynamic influence, like the eigenvalue centrality, is the proportion of infinite walks starting from each node, where walk steps are scaled such that the linear dynamics of the system are expected to converge to a non-null steady state~\cite{Klemm2012}.
The impact sums, over increasing walk lengths, the probability of transmission to the end node of the walk and that the end node has not  been previously visited by a shorter walk~\cite{Bauer2012}.
Note that all of these measures are based on counts of walks.

\vspace{2em}
\stepcounter{note} \noindent
\textbf{Note S\arabic{note}. Walk counting imposes a discrete-time perspective.}
Walk counting, when understood in terms of infinte sums,
counts all walks of length $k$  before any of length $k+1$.
This may make sense for flows characterized by transfer in which the number of ``walkers'' remains constant at each time step.
It is less applicable for diffusive flow, in which the number of walkers can increase at each time step.
From an epidemiological perspective, ignoring for the moment topological constraints,
it is equally likely that 
 two successive  transmissions form
a chain $a \rightarrow b \rightarrow c$ as a star $a \rightarrow b, a \rightarrow d$. 
Yet walk-counting considers these two cases differently.

Seen from another angle, the structure of walk-counting is equivalent to imposing a discrete-time perspective on a continuous-time process.
The potential for double-exponential growth in  the  number of transmission events per discrete time step~\cite{fountoulakis2012} means that even for small $k$, a discrete-time framework induces substantial blur.

\vspace{2em}
\stepcounter{note} \noindent
\textbf{Note S\arabic{note}. Motivation for the choice of measures for comparison.}
The predictive power of the ExF is compared  to one spreading power metric and two centrality measures. 
Spreading power metrics are a recent theme in the centrality literature specifically designed to quantify the spreading power of all nodes.
Spreading power metrics have consistently been shown to have better correlations with epidemic outcomes than previous centrality measures~\cite{daSilva2012, Klemm2012, Bauer2012}.
The accessibility is chosen to represent them as it is the most general, most powerful, and most theoretically developed published spreading power metric. 
In contrast, the authors of the dynamic influence write that their metric is only accurate when the actual transmission probability is close to the assumed optimal value contained in their method \cite{Klemm2012}. 
The impact  is not defined for SI models, has a different definition for SIS than for SIR models, and directly depends on the transmission probability of a specific spreading process \cite{Bauer2012}.

Given that almost all centrality measures are based on sums of walks of various lengths~\cite{Borgatti2005,Borgatti2006,Estrada2010, Benzi2013}, and that the best choice for walk length is influenced by network topology \cite{Borgatti2006,Bauer2012,Sikic2013},
we here make comparison to the two measures which represent the endpoints of the spectrum (see Note S4).
The eigenvalue centrality counts the number of infinite walks, and degree counts walks of length one.
We here compare to the k-shell instead of degree, as the k-shell can be thought of as the minimal degree of a node~\cite{Seidman1983}, and is a better measure of node influence~\cite{kitsak2010}.
No comparison is made to subgraph centrality as it is computationally too 
expensive to compute for large graphs \cite{igraph}.
No comparison is made to pagerank because its rankings are highly dependent on the damping factor, including frequent rank reversals after small changes~\cite{Bressan2010,Son2012}, and as it is also unstable in the face of small pertubations in network topology \cite{Ghoshal2011}.

The eigenvalue centrality and the k-shell are computed using the Igraph package version 0.6.5~\cite{igraph}.
We implement the latest version of the accessibility \cite{Viana2012} in R, following python code supplied by the authors and using sparse matrix multiplication from the Matrix package version 1.0-10 ~\cite{Bates2012}.

\vspace{2em}
\stepcounter{note} \noindent
\textbf{Note S\arabic{note}. Centrality measures exist on a spectrum}
The concept of centrality as an infinite sum of walks is found as early as Bonacich's 1987 ``Power and Centrality: a Family of Measures'' \cite{Bonacich1987}, which proposes a parameterized family of centralities
\begin{equation}
c(\beta)=\sum_{k=0}^\infty \beta^k A^{\kappa + 1}
\end{equation}
where $A$ is the network adjacency matrix and $k$ is walk length.
Bonacich explicitly states that $\beta$ controls the weighting of local vs global structure, alternately, that $(1-\beta)^{-1}$ is the radius of interest (when $\beta > 0$). This measure becomes the eigenvalue centrality when $\beta$ approaches the inverse of the largest eigenvalue of the adjacency matrix, and the degree when $\beta$ approaches zero. Bonacich concludes that the optimal choice of $\beta$ depends on the nature of communication and authority in the network~\cite{Bonacich1987}.

More recently, Estrada has proposed a generalized version of the subgraph centrality which allows the user to set the balance between local and global scale~\cite{Estrada2010}.
The subgraph centrality counts the number of closed walks for a given node (the number of walks both begining and ending at that node). These are given for every node by the trace of the exponential of the adjacency matrix. Noting that
\begin{equation}
\exp(A)=I + A + \frac{A^2}{2!} + \frac{A^3}{3!} + \ldots + \frac{A^k}{k!} + \ldots
\end{equation}
the subgraph centrality of node $i$ is
\begin{equation}
C_{subgraph}(i)=\sum_{k=0}^\infty \frac{u_k(i)}{k!}
\end{equation}
where $u_k(i)$ is the number of closed walks of length $k$ starting from node $i$.
The generalized subgraph centrality is created by shifting the factorial denominators forward or backward by $t$ steps,
\begin{equation}
C_{subgraph}^t(i)=\sum_{k=0}^\infty \frac{u_k(i)}{(k+t)!}
\end{equation}
Positive $t$ places more emphasis on local structure,
negative $t$ on global (with the understanding that the factorial of a negative number is 1)~\cite{Estrada2010}.

Benzi and Klymko unite the generalizations  proposed by Bonacich and Estrada~\cite{Benzi2013}.
Rescaling the adjacency matrix allows many centralities (i.e.  eigenvalue, Katz,  total communicability) to be expressed in Bonacich's format
\begin{equation}
\sum_{k=0}^\infty \beta^kA_R^k
\end{equation}
where $A_R$ is the appropriately rescaled adjacency matrix.
Others, i.e. the subgraph centrality, resolvent subgraph centrality, etc., are expressed in terms of the 
exponential of the (potentially rescaled) adjacency matrix, which has the following power series expansion
\begin{equation}
\sum_{k=0}^\infty \frac{(\beta A_R)^k}{k!}
\end{equation}
Benzi and Klymko further show that both formulas imply similar behavior as the tuning parameter $\beta$ approaches its limit, regardless of the rescaling of the adjacency matrix.
As $|\beta|$ goes to zero, both formulations converges towards node degree, while as it approaches its upper limit ($1/\lambda$ in the first case, where $\lambda$ is the largest eigenvalue, and $\infty$ in the second), the series converges towards the eigenvalue centrality. 
The rate of convergence depends on the spectral gap of the adjacency matrix, with a larger gap corresponding to faster convergence~\cite{Benzi2013}.

\vspace{2em}
\stepcounter{note} \noindent
\textbf{Note S\arabic{note}. Correlation between $\mathbf{ExF_2}$ and $\mathbf{ExF_3}$.}
The ExF is based on the distribution of the force of infection after an arbitrary number of infection events; a subscript can be used to indicate the number of events considered. 
Evidence that two events is sufficient is provided by the tight correlation between the metric when computed using two and again with three events ($ExF_2$ and $ExF_3$) for the simulated network classes considered here.
The mean and standard deviations in the correlations, taken over 50 networks in each class, are as follows:
Pareto  0.96 $\pm$ 0.007,
Amazon  0.95 $\pm$ 0.013,
Internet  0.97 $\pm$ 0.007, 
Facebook 0.99 $\pm$ 0.014,
Astrophysics  0.99  $\pm$ 0.005.
As expected from these tight correlations, increasing the number of events to three does not provide any meaningful increase in predictive accuracy.

\vspace{2em}
\stepcounter{note} \noindent
\textbf{Note S\arabic{note}. Invariance of  $\mathbf{ExF^M}$ to choice of scaling parameter.}
The modified version of the ExF is defined in the main text as:
\begin{displaymath}
ExF^{M}(i)= \log(\alpha\, deg(i))\, ExF(i)
\end{displaymath}
where the degree of the node is scaled by $\alpha$ so as to prevent the logarithm from being zero for nodes with degree one. 
A simple shuffling of terms clarifies the influence of $\alpha$: $ExF^M= \log(\alpha) ExF(i) + \log(deg(i)) ExF(i)$, 
implying that as $\alpha \rightarrow 1$, the scaling factor becomes irrelevant, and as $\alpha \rightarrow \infty$, it eclipses any contribution from the degree.
The manuscript suggest $\alpha=2$ is a reasonable choice, providing the needed scaling without unduly skewing the measure.

We here show that the measure is largely invariant to the choice of $\alpha$ by testing the following values:
  1.0001,  1.001,  1.01,  1.1,  1.5, 2, 3, 4, 8, and 16.
For each $\alpha$ tested, 
the correlation between $ExF^M$ at $\alpha=2$ and the test value  is measured.
Measurements are made on all non-hub nodes for one hundred networks of each of the five simulated network families. 
The mean value for each parameter/network type is reported in Table S1.
We test over the full network as this is likely to bias the testing values towards low degree nodes, where the choice of $\alpha$ is more likely to have an effect.

All correlations are greater than 0.999 for $\alpha$ in the range 1.5--3.
Only two cases show correlation less than 0.99, both occurring when $\alpha=16$. 
If instead of reporting the mean correlation observed over the one hundred networks, we report the minimum, the same patterns hold, with the lowest value dropping to 0.976, again for $\alpha=16$.

\begin{table}
\textbf{Table S1. Correlation between $\mathbf{ExF^M}$ computed with scaling parameter 2 and with the value given in the column headings.}
\begin{ruledtabular}
\begin{tabular}{rrrrrr}
 & 1.0001 & 1.001 & 1.01 & 1.1 & 1.5 \\ 
  \hline
Pareto & 0.9958 & 0.9959 & 0.9960 & 0.9971 & 0.9994 \\ 
  Amazon & 0.9980 & 0.9980 & 0.9981 & 0.9986 & 0.9997 \\ 
  Internet & 0.9965 & 0.9965 & 0.9966 & 0.9975 & 0.9995 \\ 
  Facebook & 0.9995 & 0.9995 & 0.9995 & 0.9996 & 0.9999 \\ 
  Astrophysics & 0.9996 & 0.9996 & 0.9996 & 0.9997 & 0.9999 \\ 
   \hline
 \hline
 & 2 & 3 & 4 & 8 & 16 \\ 
  \hline
Pareto & 1.0000 & 0.9992 & 0.9980 & 0.9942 & 0.9899 \\ 
  Amazon & 1.0000 & 0.9996 & 0.9989 & 0.9966 & 0.9939 \\ 
  Internet & 1.0000 & 0.9992 & 0.9979 & 0.9931 & 0.9873 \\ 
  Facebook & 1.0000 & 0.9999 & 0.9996 & 0.9987 & 0.9975 \\ 
  Astrophysics & 1.0000 & 0.9999 & 0.9997 & 0.9989 & 0.9979 \\ 
\end{tabular}
\end{ruledtabular}
\end{table}

\vspace{2em}
\stepcounter{note} \noindent
\textbf{Note S\arabic{note}. Running times.}
Straightforward calculation of the ExF has time complexity $O(n_1^2 * n_2)$, where $n_1$ and $n_2$ are the number of neighbors at geodesic distance one and two from the seed. 
It is difficult to analytically compare a time complexity computed on individual nodes with time complexities whose calculation is based on the entire adjacency matrix.
Simple benchmarking experiments, however, suggest that the ExF scales approximately linearly in the number of edges and nodes in the network.
We first compare the median running time of all measures on fifty Pareto networks of 1,000 nodes, with running time measured at sub-microsecond accuracy~\cite{mb2013} as the median computation time over ten runs on each network.
The k-shell is computed in 0.5\% of the time,
and the eigenvalue centrality in 11\% of the time needed to compute the ExF for all non-hub nodes.
Computing the accessibility takes 84 times longer than the ExF. 
The benchmarking is then repeated with the same protocol on 10,000 node Pareto networks.
Increases in running time for the k-shell (7x), eigenvalue centrality (9x), and ExF (16x) have roughly linear correspondence to the tenfold increase in the number of network nodes.
Recall that 
the proven time complexity for the k-shell and expected time for the eigenvalue centrality are both $O(|V|+|E|)$, i.e. linear.
As expected, the accessibility does not scale well, with a ten-fold increase in network size leading to a 265 fold increase in median running time.
Recall that it is computed by taking powers of the adjacency matrix, i.e. something worse than $O(|V|^{2.4})$.
Computation of k-shell and eigenvalue centrality takes advantage of the highly optimized code implemented in the Igraph package \cite{igraph}.
The accessibility is computed in native R~\cite{R} code using sparse matrix multiplication from the Matrix package 1.0-10 ~\cite{Bates2012}.
The ExF is computed in C++ code via an R interface.

\vspace{2em}
\stepcounter{note} \noindent
\textbf{Note S\arabic{note}. Agreement between the ExF, k-shell, and eigenvalue centrality on the most important nodes.}
We here assess the agreement between  the ExF, k-shell, and eigenvalue centrality as to which nodes are the most important in the network.
All three measures are compared on one hundred networks for each of the five families. 
Table S2 shows the mean rank correlation between ExF and the other measures, as well as the agreement between ExF and eigenvalue centrality regarding the top ten nodes.

Overlap with the k-shell is problematic
 in that the k-shell does not provide deep resolution. 
In the looser networks, the highest k-shell contains a large percentage of the total nodes in the network (mean 41\% in Pareto, 92\% in Amazon). 
Even in the denser networks, the top k-shell contains more than 10\% of the network nodes (Internet 15\%, Facebook 13\%, Astrophysics 14\%).
Hence the observation that the top 10 nodes (1\% of the network) as ranked by the ExF are also found in the highest k-shell is not sufficiently meaningful to report in the table.

\begin{table}
\textbf{Table S2. Pearson's rank correlation coefficient between the ExF and the eigenvalue centrality (ev) and k-shell (ks), along with the mean number of nodes placed in the top ten by both $\mathbf{ExF}$ and eigenvalue centrality, by network family.}
\begin{ruledtabular}
\begin{tabular}{rrrr}
 & $\rho$-ev & $\rho$-ks & overlap \\ 
  \hline
Pareto & 0.62 & 0.79 & 6.38 \\ 
  Amazon & 0.71 & 0.61 & 6.96 \\ 
  Internet & 0.73 & 0.82 & 9.16 \\ 
  Facebook & 0.83 & 0.90 & 9.35 \\ 
  Astrophysics & 0.84 & 0.92 & 9.22 \\ 
\end{tabular}
\end{ruledtabular}
\end{table}

\vspace{2em}
\stepcounter{note} \noindent
\textbf{Note S\arabic{note}. Sufficiency of local neighborhood.}
Reliance on a local neighborhood is consistent with established theory showing that topological information content falls off quickly with distance.
Bonacich demonstrated in 1987 that the influence of walks must decay at least exponentially in walk length to guarantee convergence~\cite{Bonacich1987}. 
More recent work shows that even faster decay rates are often motivated~\cite{Estrada2005,Estrada2010,Benzi2013}.
The fall-off in information can also be shown by the following example.
Consider a long linear chain of nodes which ultimately connects to a network hub. Let  $\beta$ be the transmission/recovery ratio in a process with recovery and $\Delta_i$ the distance from the $i^{th}$ node of the chain to the hub. 
If the spreading process reaches the hub, an epidemic is almost certain. The probability of this occuring is at best $\beta^{\Delta_i}$.
For $\beta<0.1$, this probability is estimatable to three of four decimal places using only local information.
More generally, since epidemic spread is almost instantaneous on scale free networks~\cite{lloyd2001,fountoulakis2012}, the expectation is that the time-step which takes a process outside the local neighborhood of its origin brings it to the majority of the network.

\vspace{2em}
\stepcounter{note} \noindent
\textbf{Note S\arabic{note}. Selection of seed nodes.}
For each simulated network, the ExF is computed for all non-hub nodes, with hubs defined as nodes whose degree is greater than 60\% of the maximum degree node in the network.
The range of observed ExF values is split into 15 equal width bins.
Five seed nodes are selected uniformly at random from each bin, giving approximately 75 seed nodes for each of the hundred simulated networks in each of the five network families.
The reason for the binning is that in scale free networks, the distribution of any centrality measure is strongly biased towards low values. Selecting seed nodes across the full range of the ExF ensures that nodes with high and medium ExF are also included in all tests.
That hubs have maximal spreading power is already firmly estalished, and also the reasons why; they also have the highest ExF values. Further, under this definition, less than 0.1\% of nodes are hubs. 

\vspace{2em}
\stepcounter{note} \noindent
\textbf{Note S\arabic{note}. Probability of epidemic.}
This work defines the epidemic outcome in SIS/SIR processes as the probability that an epidemic occurs.
This is in contrast to the measure typically used, the mean number of infected nodes (i.e.~\cite{kitsak2010,Sikic2013,daSilva2012,Viana2012,Bauer2012,Klemm2012,Newman2002}).
We are not convinced that the mean is a good summary statistic. 
In over 20,000 simulated continuous-time SIS spreading processes, 
no processes which went extinct reached more than 20 nodes, while
processes which did not go extinct reached the majority of the network.
It has been argued that such bifurcation in outcomes is predicted by theory~\cite{Wilkinson2013}.
Given that the distribution of the number of infected nodes is characterized by two well separated modes,
the mean is best seen as an indirect estimate of the likelihood of the higher mode. It is this likelihood which we directly measure as the epidemic potential.

\newpage
\begin{table}


\textbf{Table S3. Empirical evidence that fixed multiples of the inverse of the largest eigenvalue give $\beta$ in the critical range for epidemic take-off.}
Percentage of nodes with epidemic potential in the interval $[2\%, 98\%]$ when $\beta$ is set to the given multiple of the $1/\lambda$ (where $\lambda$ is the largest eigenvalue of the adjacency matrix), by network type. 
Epidemic potential is measured by simulating 100 spreading processes seeded from the given node and counting how many result in an epidemic. A different multiple is used for each type of spreading process. Denser networks require higher multiples. 

\noindent
\begin{ruledtabular}
\begin{tabular}{lrcrc}
      & \multicolumn{2}{c}{Pareto/Amazon}& 
        \multicolumn{2}{c}{Internet/Astro/Facebook}\\
      \hline
      & multiple & percentage      & multiple & percentage\\
SIS-C & 4.0 & (100/100)\% & 8.0 & (99/99/98) \\ 
SIS-D & 2.0 & (93/89)\%   & 2.5 & (83/81/77)\\
SIR-C & 6.0 & (100/100)\% & 8.0 & (99/98/96) \\
SIR-D & 3.5 & (83/76)\%   & 5.0 & (76/78/81) \\
\end{tabular}
\end{ruledtabular}
\end{table}

\begin{table}[b]
 \textbf{Table S4. Simulated network families.}
 The  mean diameter, mean graph density, and
 empirical middle 65\% quantile range of the largest eigenvalue for the different network families.
Pareto and Amazon co-purchase networks have a large, loose structure with low eigenvalue, suggesting less inherent susceptibility to epidemics than the smaller and more dense collaboration networks; Google's map of the Internet lies in between.
Means and standard deviations are computed over 100 simulated networks with 1,000 nodes.
\begin{ruledtabular}
\begin{tabular}{llll}
                                 & diameter       & density  & 65\% quantile \\
Pareto                           & 11.6 $\pm 1.0$ & 3.2 e-04 &  7.1 -- 10.1 \\  
Amazon   \cite{Leskovec2007a}    &  7.2 $\pm 0.4$ & 6.9 e-04 & 10.1 -- 13.7 \\
Internet   \cite{Leskovec2007a}    &  7.0 $\pm 0.5$ & 9.4 e-03 & 25.2 -- 35.2 \\
Astrophysics \cite{Leskovec2007} &  5.5 $\pm 0.6$ & 2.1 e-02 & 54.5 -- 61.9 \\
Facebook \cite{viswanath2009}    &  5.5 $\pm 0.5$ & 2.4 e-02 & 65.2 -- 73.7 \\
\end{tabular}
\end{ruledtabular}
\end{table}

\newpage
\clearpage
\begin{table}[h]
 \textbf{Table S5. Empirical networks.}
The number of nodes, 90$^{th}$ percentile effective diameter, and density of the empirical networks.
Networks were downloaded from the Stanford Large Network Collection (SNAP),
Alex Arena's collection (AA), and the Max Planck Institute for Software Systems website (MPI), which in turn credit the cited publication for the network.
\begin{ruledtabular}
\begin{tabular}{lrrrr}
            & nodes & diameter & density  & source \\
PGPgiantcompo & 10680 & 10.0 & 4.26 e-4 &AA \cite{Boguna2004} \\
amazon0302   & 262111 & 11.1 & 0.26 e-4 & SNAP \cite{Leskovec2007a}\\
amazon0601    & 403364 & 7.6 & 0.30 e-4 & SNAP \cite{Leskovec2007a}\\
ca-AstroPh     & 17903 & 5.0 & 12.30 e-4& SNAP \cite{Leskovec2007}\\
ca-CondMat   & 21363 & 6.5 & 4.01  e-4 & SNAP \cite{Leskovec2007}\\
ca-GrQc       & 4158 & 7.6 & 15.53  e-4 & SNAP \cite{Leskovec2007}\\
ca-HepPh & 11204 & 5.8     & 18.74 e-4  & SNAP \cite{Leskovec2007}\\
ca-HepTh       & 8638 & 7.4 & 6.65  e-4 & SNAP \cite{Leskovec2007}\\
cit-HepPh & 34401 & 5.0 & 7.11  e-4 & SNAP \cite{Leskovec2005}\\
cit-HepTh & 27400 & 5.3 & 9.38  e-4 & SNAP \cite{Leskovec2005}\\
com-dblp & 317080 & 8.0 & 0.21  e-4 &SNAP \cite{Yang2012}\\
email-EuAll & 224832 & 4.5 & 0.13  e-4 & SNAP \cite{Leskovec2007}\\
email-Uni   & 1133 & 4.3 & 85.00  e-4 & AA \cite{Guimera2003}\\
facebooklcc & 59691 & 5.6 & 4.09  e-4 & MPI \cite{viswanath2009}\\
loc-brightkite & 56739 & 6.0 & 1.32  e-4 & SNAP \cite{Cho2011}\\
loc-gowalla & 196591 & 5.7 & 0.49  e-4 & SNAP \cite{Cho2011}\\
p2p-Gnutella31 & 62561 & 6.7 & 0.76  e-4 & SNAP \cite{Leskovec2007}\\
soc-Epinions1 & 75877 & 5.0 & 1.41  e-4  & SNAP \cite{Richardson2003}\\
soc-Slashdot0902 & 82168 & 4.7 & 1.49  e-4 & SNAP \cite{Leskovec2009}\\
soc-sign-epinions & 119130 & 4.9 & 0.99  e-4 & SNAP \cite{Leskovec2010}\\
web-Google & 855802 & 8.1 & 0.12  e-4 & SNAP \cite{Leskovec2009}\\
web-NotreDame & 325729 & 9.4 & 0.21  e-4 & SNAP \cite{Albert1999}\\
web-Stanford & 255265 & 9.7 & 0.60  e-4 & SNAP \cite{Leskovec2009}\\
wiki-Vote & 7066 & 3.8 & 40.36  e-4 & SNAP \cite{Leskovec2010}\\
\end{tabular}
\end{ruledtabular}

\begin{tabular}{ll}
SNAP & http://snap.stanford.edu/data/index.html\\
AA & http://deim.urv.cat/~aarenas/data/welcome.htm\\
MPI & http://socialnetworks.mpi-sws.org/data-wosn2009.html\\
\end{tabular}

\end{table}

\newpage
\bibliographystyle{pnas}
\bibliography{cleanbib}
\vspace{2em}
\textbf{Results tables on following pages}
\clearpage

\onecolumngrid

\pagebreak
\setcounter{table}{5}
\stepcounter{table}
\noindent
 \textbf{Table S\arabic{table}. Mean correlations between node spreading power metrics and epidemic outcomes on each type of spreading process on the simulated networks, by network model.}
Shown is the mean and standard error in correlations measured on one hundred networks from each family. This information is duplicated in Figure 2 in the main text. ExF$^M$ is not included for SI models as the modification only makes sense for processes with recovery.
Spreading processes are suffixed to indicate simulations in continuous (-C) or discrete (-D) time. Epidemic outcomes are time to half coverage for SI processes and epidemic potential in the remaining processes.

\vspace{\topsep}
\noindent
\begin{tabular}{lcccccccccc}
&ExF & ExF$^M$ & accesibility & eigenvalue cent. & k-shell\\
  \hline  
Pareto\\
  \hline \\ [-1.8ex]  
SI  & 0.84 $\pm$ 0.04 & & 0.66 $\pm$ 0.05 & 0.38 $\pm$ 0.06 & 0.76 $\pm$ 0.05\\
SIS-C  & 0.93 $\pm$ 0.02 & 0.89 $\pm$ 0.02 & 0.78 $\pm$ 0.05 & 0.53 $\pm$ 0.08 & 0.77 $\pm$ 0.05\\
SIS-D  & 0.94 $\pm$ 0.02 & 0.91 $\pm$ 0.02 & 0.78 $\pm$ 0.06 & 0.51 $\pm$ 0.09 & 0.79 $\pm$ 0.05\\
SIR-C  & 0.91 $\pm$ 0.02 & 0.81 $\pm$ 0.03 & 0.71 $\pm$ 0.05 & 0.43 $\pm$ 0.08 & 0.82 $\pm$ 0.03\\
SIR-D  & 0.87 $\pm$ 0.14 & 0.78 $\pm$ 0.13 & 0.68 $\pm$ 0.12 & 0.40 $\pm$ 0.09 & 0.82 $\pm$ 0.14\\
  \hline  
Amazon\\
  \hline \\ [-1.8ex]  
SI    & 0.87 $\pm$ 0.02 &                 & 0.84 $\pm$ 0.02 & 0.44 $\pm$ 0.05 & 0.88 $\pm$ 0.02\\
SIS-C  & 0.95 $\pm$ 0.01 & 0.92 $\pm$ 0.01 & 0.91 $\pm$ 0.02 & 0.63 $\pm$ 0.06 & 0.72 $\pm$ 0.05\\
SIS-D  & 0.95 $\pm$ 0.01 & 0.91 $\pm$ 0.02 & 0.92 $\pm$ 0.03 & 0.59 $\pm$ 0.06 & 0.74 $\pm$ 0.06\\
SIR-C  & 0.92 $\pm$ 0.02 & 0.85 $\pm$ 0.02 & 0.88 $\pm$ 0.03 & 0.53 $\pm$ 0.05 & 0.82 $\pm$ 0.04\\
SIR-D  & 0.90 $\pm$ 0.02 & 0.81 $\pm$ 0.03 & 0.87 $\pm$ 0.04 & 0.46 $\pm$ 0.06 & 0.87 $\pm$ 0.04\\
  \hline  
Internet\\
  \hline \\ [-1.8ex]  
SI    & 0.82 $\pm$ 0.03 &                 & 0.77 $\pm$ 0.08 & 0.37 $\pm$ 0.04 & 0.73 $\pm$ 0.04\\
SIS-C  & 0.92 $\pm$ 0.03 & 0.90 $\pm$ 0.01 & 0.61 $\pm$ 0.09 & 0.65 $\pm$ 0.04 & 0.95 $\pm$ 0.01\\
SIS-D  & 0.85 $\pm$ 0.03 & 0.96 $\pm$ 0.01 & 0.45 $\pm$ 0.08 & 0.82 $\pm$ 0.04 & 0.89 $\pm$ 0.03\\
SIR-C  & 0.92 $\pm$ 0.02 & 0.90 $\pm$ 0.01 & 0.62 $\pm$ 0.09 & 0.66 $\pm$ 0.04 & 0.95 $\pm$ 0.01\\
SIR-D  & 0.89 $\pm$ 0.03 & 0.87 $\pm$ 0.02 & 0.60 $\pm$ 0.09 & 0.60 $\pm$ 0.04 & 0.98 $\pm$ 0.01\\
 \hline  
Astrophysics\\
 \hline \\ [-1.8ex]  
SI  & 0.81 $\pm$ 0.02 & & 0.51 $\pm$ 0.07 & 0.36 $\pm$ 0.03 & 0.6 $\pm$ 0.04\\
SIS-C  & 0.92 $\pm$ 0.01 & 0.94 $\pm$ 0.01 & 0.31 $\pm$ 0.05 & 0.71 $\pm$ 0.02 & 0.95 $\pm$ 0.01\\
SIS-D  & 0.85 $\pm$ 0.02 & 0.97 $\pm$ 0 & 0.2 $\pm$ 0.04 & 0.86 $\pm$ 0.03 & 0.96 $\pm$ 0.01\\
SIR-C  & 0.92 $\pm$ 0.01 & 0.94 $\pm$ 0.01 & 0.31 $\pm$ 0.05 & 0.71 $\pm$ 0.02 & 0.95 $\pm$ 0.01\\
SIR-D  & 0.89 $\pm$ 0.01 & 0.92 $\pm$ 0.01 & 0.29 $\pm$ 0.05 & 0.67 $\pm$ 0.03 & 0.97 $\pm$ 0.01\\
  \hline  
Facebook\\
  \hline \\ [-1.8ex]  
SI  & 0.83 $\pm$ 0.02 & & 0.43 $\pm$ 0.1 & 0.38 $\pm$ 0.02 & 0.61 $\pm$ 0.04\\
SIS-C  & 0.9 $\pm$ 0.02 & 0.95 $\pm$ 0.01 & 0.22 $\pm$ 0.05 & 0.73 $\pm$ 0.02 & 0.95 $\pm$ 0.01\\
SIS-D  & 0.82 $\pm$ 0.02 & 0.97 $\pm$ 0 & 0.14 $\pm$ 0.04 & 0.87 $\pm$ 0.02 & 0.97 $\pm$ 0.01\\
SIR-C  & 0.9 $\pm$ 0.02 & 0.95 $\pm$ 0.01 & 0.22 $\pm$ 0.05 & 0.73 $\pm$ 0.02 & 0.95 $\pm$ 0.01\\
SIR-D  & 0.87 $\pm$ 0.02 & 0.94 $\pm$ 0.01 & 0.2 $\pm$ 0.05 & 0.7 $\pm$ 0.03 & 0.97 $\pm$ 0.01\\
\end{tabular}

\pagebreak
\stepcounter{table}
\noindent
 \textbf{Table S\arabic{table}. Correlation between spreading power metrics and \emph{tthc} in real world networks.}
Shown is the estimated correlation from 1,000 nodes on the given network, along with the 95\% confidence bounds of the estimate.
This information is duplicated in Figure 3 in the main text. The ExF$^M$ is not included here as the modification only makes sense for processes with recovery; an empty column is used to allow easier visual comparison with the remaining tables. Accessibility is not measured for networks with more than 25,000 nodes.

\vspace{\topsep}
\noindent
\begin{tabular}{lccccc} 
&ExF & ExF$^M$ & accessibility & eigenvalue cent. & k-shell\\
\hline \\ [-1.3ex]
PGPgiantcompo & 0.69 $\pm$ 0.03 &     --         & 0.58 $\pm$ 0.04 & 0.19 $\pm$ 0.06 & 0.43 $\pm$ 0.05 \\
amazon0302 & 0.54 $\pm$ 0.04 &     --         &     --         & 0.15 $\pm$ 0.06 & 0.30 $\pm$ 0.06 \\
amazon0601 & 0.74 $\pm$ 0.03 &     --         &     --         & 0.09 $\pm$ 0.06 & 0.63 $\pm$ 0.04 \\
ca-AstroPh & 0.84 $\pm$ 0.02 &     --         & 0.49 $\pm$ 0.05 & 0.29 $\pm$ 0.06 & 0.58 $\pm$ 0.04 \\
ca-CondMat & 0.84 $\pm$ 0.02 &     --         & 0.53 $\pm$ 0.04 & 0.26 $\pm$ 0.06 & 0.65 $\pm$ 0.04 \\
ca-GrQc & 0.78 $\pm$ 0.02 &     --         & 0.58 $\pm$ 0.04 & 0.16 $\pm$ 0.06 & 0.36 $\pm$ 0.05 \\
ca-HepPh & 0.82 $\pm$ 0.02 &     --         & 0.54 $\pm$ 0.04 & 0.20 $\pm$ 0.06 & 0.39 $\pm$ 0.05 \\
ca-HepTh & 0.78 $\pm$ 0.02 &     --         & 0.56 $\pm$ 0.04 & 0.05 $\pm$ 0.06 & 0.47 $\pm$ 0.05 \\
cit-HepPh & 0.82 $\pm$ 0.02 &     --         &     --         & 0.28 $\pm$ 0.06 & 0.68 $\pm$ 0.03 \\
cit-HepTh & 0.84 $\pm$ 0.02 &     --         & 0.57 $\pm$ 0.04 & 0.38 $\pm$ 0.05 & 0.64 $\pm$ 0.04 \\
com-dblp & 0.79 $\pm$ 0.02 &     --         &     --         & 0.05 $\pm$ 0.06 & 0.36 $\pm$ 0.05 \\
email-EuAll & 0.41 $\pm$ 0.05 &     --         &     --         & 0.34 $\pm$ 0.05 & 0.50 $\pm$ 0.05 \\
email-Uni & 0.92 $\pm$ 0.01 &     --         & 0.61 $\pm$ 0.04 & 0.56 $\pm$ 0.04 & 0.84 $\pm$ 0.02 \\
facebooklcc & 0.86 $\pm$ 0.02 &     --         &     --         & 0.19 $\pm$ 0.06 & 0.59 $\pm$ 0.04 \\
loc-brightkite & 0.79 $\pm$ 0.02 &     --         &     --         & 0.13 $\pm$ 0.06 & 0.54 $\pm$ 0.04 \\
loc-gowalla & 0.66 $\pm$ 0.03 &     --         &     --         & 0.25 $\pm$ 0.06 & 0.53 $\pm$ 0.04 \\
p2p-Gnutella31 & 0.94 $\pm$ 0.01 &     --         & 0.72 $\pm$ 0.03 & 0.53 $\pm$ 0.04 & 0.92 $\pm$ 0.01 \\
soc-Epinions1 & 0.80 $\pm$ 0.02 &     --         &     --         & 0.33 $\pm$ 0.06 & 0.47 $\pm$ 0.05 \\
soc-Slashdot0902 & 0.84 $\pm$ 0.02 &     --         &     --         & 0.42 $\pm$ 0.05 & 0.60 $\pm$ 0.04 \\
soc-sign-epinions & 0.81 $\pm$ 0.02 &     --         &     --         & 0.29 $\pm$ 0.06 & 0.47 $\pm$ 0.05 \\
web-Google & 0.69 $\pm$ 0.02 &     --         &     --         & 0.07 $\pm$ 0.04 & 0.59 $\pm$ 0.02 \\
web-NotreDame & 0.43 $\pm$ 0.05 &     --         &     --         & 0.18 $\pm$ 0.06 & 0.26 $\pm$ 0.06 \\
web-Stanford & 0.25 $\pm$ 0.06 &     --         &     --         & 0.06 $\pm$ 0.06 & 0.12 $\pm$ 0.06 \\
wiki-Vote & 0.86 $\pm$ 0.02 &     --         & 0.50 $\pm$ 0.05 & 0.50 $\pm$ 0.05 & 0.72 $\pm$ 0.03 \\
\end{tabular} 

\pagebreak
\stepcounter{table}
\noindent
 \textbf{Table S\arabic{table}. Correlation between spreading power metrics and epidemic potential in discrete time SIS processes on real world networks.}
Shown is the estimated correlation from 1,000 nodes on the given network, along with the 95\% confidence bounds of the estimate. This information is duplicated in Figure 3 in the main text. Accessibility is not measured for networks with more than 25,000 nodes.

\vspace{\topsep}
\noindent
\begin{tabular}{lccccc} 
&ExF & ExF$^M$ & accessibility & eigenvalue cent. & k-shell\\
\hline \\ [-1.3ex]
PGPgiantcompo & 0.87 $\pm$ 0.02 & 0.91 $\pm$ 0.01 & 0.62 $\pm$ 0.04 & 0.33 $\pm$ 0.06 & 0.77 $\pm$ 0.03 \\
amazon0302 & 0.79 $\pm$ 0.02 & 0.80 $\pm$ 0.02 &     --         & 0.12 $\pm$ 0.06 & 0.51 $\pm$ 0.05 \\
amazon0601 & 0.77 $\pm$ 0.03 & 0.87 $\pm$ 0.02 &     --         & -0.01 $\pm$ 0.06 & 0.68 $\pm$ 0.03 \\
ca-AstroPh & 0.94 $\pm$ 0.01 & 0.92 $\pm$ 0.01 & 0.51 $\pm$ 0.05 & 0.41 $\pm$ 0.05 & 0.83 $\pm$ 0.02 \\
ca-CondMat & 0.93 $\pm$ 0.01 & 0.92 $\pm$ 0.01 & 0.59 $\pm$ 0.04 & 0.35 $\pm$ 0.05 & 0.85 $\pm$ 0.02 \\
ca-GrQc & 0.92 $\pm$ 0.01 & 0.92 $\pm$ 0.01 & 0.52 $\pm$ 0.05 & 0.25 $\pm$ 0.06 & 0.7 $\pm$ 0.03 \\
ca-HepPh & 0.92 $\pm$ 0.01 & 0.90 $\pm$ 0.01 & 0.51 $\pm$ 0.05 & 0.28 $\pm$ 0.06 & 0.55 $\pm$ 0.04 \\
ca-HepTh & 0.92 $\pm$ 0.01 & 0.93 $\pm$ 0.01 & 0.64 $\pm$ 0.04 & 0.10 $\pm$ 0.06 & 0.72 $\pm$ 0.03 \\
cit-HepPh & 0.93 $\pm$ 0.01 & 0.94 $\pm$ 0.01 & 0.55 $\pm$ 0.04 & 0.38 $\pm$ 0.05 & 0.93 $\pm$ 0.01 \\
cit-HepTh & 0.93 $\pm$ 0.01 & 0.93 $\pm$ 0.01 & 0.71 $\pm$ 0.03 & 0.57 $\pm$ 0.04 & 0.90 $\pm$ 0.01 \\
com-dblp & 0.90 $\pm$ 0.01 & 0.55 $\pm$ 0.03 &     --         & 0.08 $\pm$ 0.04 & 0.55 $\pm$ 0.03 \\
email-EuAll & 0.36 $\pm$ 0.05 & 0.9 $\pm$ 0.01 &     --         & 0.64 $\pm$ 0.04 & 0.85 $\pm$ 0.02 \\
email-Uni & 0.95 $\pm$ 0.01 & 0.97 $\pm$ 0.00 & 0.61 $\pm$ 0.04 & 0.75 $\pm$ 0.03 & 0.97 $\pm$ 0.00 \\
facebooklcc & 0.93 $\pm$ 0.01 & 0.94 $\pm$ 0.01 &     --         & 0.31 $\pm$ 0.06 & 0.88 $\pm$ 0.01 \\
loc-brightkite & 0.85 $\pm$ 0.02 & 0.95 $\pm$ 0.01 & 0.58 $\pm$ 0.04 & 0.29 $\pm$ 0.06 & 0.85 $\pm$ 0.02 \\
loc-gowalla & 0.68 $\pm$ 0.03 & 0.91 $\pm$ 0.01 &     --         & 0.51 $\pm$ 0.05 & 0.89 $\pm$ 0.01 \\
p2p-Gnutella31 & 0.95 $\pm$ 0.01 & 0.97 $\pm$ 0.00 & 0.83 $\pm$ 0.02 & 0.68 $\pm$ 0.03 & 0.92 $\pm$ 0.01 \\
soc-Epinions1 & 0.77 $\pm$ 0.03 & 0.96 $\pm$ 0.01 &     --         & 0.63 $\pm$ 0.04 & 0.85 $\pm$ 0.02 \\
soc-Slashdot0902 & 0.80 $\pm$ 0.02 & 0.97 $\pm$ 0.00 &     --         & 0.71 $\pm$ 0.03 & 0.93 $\pm$ 0.01 \\
soc-sign-epinions & 0.76 $\pm$ 0.03 & 0.96 $\pm$ 0.01 &     --         & 0.54 $\pm$ 0.04 & 0.81 $\pm$ 0.02 \\
web-Google & 0.79 $\pm$ 0.02 & 0.93 $\pm$ 0.01 &     --         & 0.10 $\pm$ 0.06 & 0.91 $\pm$ 0.01 \\
web-NotreDame & 0.73 $\pm$ 0.03 & 0.92 $\pm$ 0.01 &     --         & 0.34 $\pm$ 0.06 & 0.49 $\pm$ 0.05 \\
web-Stanford & 0.70 $\pm$ 0.03 & 0.89 $\pm$ 0.01 &     --         & 0.38 $\pm$ 0.05 & 0.77 $\pm$ 0.03 \\
wiki-Vote & 0.94 $\pm$ 0.01 & 0.96 $\pm$ 0.01 & 0.48 $\pm$ 0.05 & 0.71 $\pm$ 0.03 & 0.95 $\pm$ 0.01 \\
\end{tabular} 

\pagebreak
\stepcounter{table}
\noindent
\textbf{Table S\arabic{table}. Correlation between spreading power metrics and epidemic potential in discrete time SIR processes on real world networks.}
Shown is the estimated correlation from 1,000 nodes on the given network, along with the 95\% confidence bounds of the estimate. This information is duplicated in Figure 3 in the main text. Accessibility is not measured for networks with more than 25,000 nodes.

\vspace{\topsep}
\noindent
\begin{tabular}{lccccc} 
&ExF & ExF$^M$ & accessibility & eigenvalue cent. & k-shell\\
\hline \\ [-1.3ex]
PGPgiantcompo & 0.82 $\pm$ 0.02 & 0.86 $\pm$ 0.02 & 0.69 $\pm$ 0.03 & 0.36 $\pm$ 0.05 & 0.73 $\pm$ 0.03 \\
amazon0302 & 0.67 $\pm$ 0.03 & 0.65 $\pm$ 0.04 &     --         & 0.19 $\pm$ 0.06 & 0.33 $\pm$ 0.06 \\
amazon0601 & 0.75 $\pm$ 0.03 & 0.83 $\pm$ 0.02 &     --         & -0.03 $\pm$ 0.06 & 0.71 $\pm$ 0.03 \\
ca-AstroPh & 0.94 $\pm$ 0.01 & 0.93 $\pm$ 0.01 & 0.53 $\pm$ 0.04 & 0.42 $\pm$ 0.05 & 0.84 $\pm$ 0.02 \\
ca-CondMat & 0.92 $\pm$ 0.01 & 0.91 $\pm$ 0.01 & 0.65 $\pm$ 0.04 & 0.38 $\pm$ 0.05 & 0.83 $\pm$ 0.02 \\
ca-GrQc & 0.85 $\pm$ 0.02 & 0.83 $\pm$ 0.02 & 0.65 $\pm$ 0.04 & 0.27 $\pm$ 0.06 & 0.55 $\pm$ 0.04 \\
ca-HepPh & 0.91 $\pm$ 0.01 & 0.91 $\pm$ 0.01 & 0.53 $\pm$ 0.04 & 0.31 $\pm$ 0.06 & 0.59 $\pm$ 0.04 \\
ca-HepTh & 0.89 $\pm$ 0.01 & 0.88 $\pm$ 0.01 & 0.70 $\pm$ 0.03 & 0.10 $\pm$ 0.06 & 0.64 $\pm$ 0.04 \\
cit-HepPh & 0.93 $\pm$ 0.01 & 0.94 $\pm$ 0.01 & 0.56 $\pm$ 0.04 & 0.39 $\pm$ 0.05 & 0.94 $\pm$ 0.01 \\
cit-HepTh & 0.93 $\pm$ 0.01 & 0.92 $\pm$ 0.01 & 0.70 $\pm$ 0.03 & 0.54 $\pm$ 0.04 & 0.87 $\pm$ 0.01 \\
com-dblp & 0.89 $\pm$ 0.01 & 0.55 $\pm$ 0.03 &     --         & 0.08 $\pm$ 0.04 & 0.53 $\pm$ 0.03 \\
email-EuAll & 0.31 $\pm$ 0.06 & 0.87 $\pm$ 0.02 &     --         & 0.60 $\pm$ 0.04 & 0.81 $\pm$ 0.02 \\
email-Uni & 0.94 $\pm$ 0.01 & 0.97 $\pm$ 0.00 & 0.63 $\pm$ 0.04 & 0.81 $\pm$ 0.02 & 0.96 $\pm$ 0.00 \\
facebooklcc & 0.93 $\pm$ 0.01 & 0.95 $\pm$ 0.01 &     --         & 0.32 $\pm$ 0.06 & 0.89 $\pm$ 0.01 \\
loc-brightkite & 0.84 $\pm$ 0.02 & 0.95 $\pm$ 0.01 & 0.61 $\pm$ 0.04 & 0.31 $\pm$ 0.06 & 0.87 $\pm$ 0.02 \\
loc-gowalla & 0.68 $\pm$ 0.03 & 0.90 $\pm$ 0.01 &     --         & 0.47 $\pm$ 0.05 & 0.86 $\pm$ 0.02 \\
p2p-Gnutella31 & 0.96 $\pm$ 0.00 & 0.97 $\pm$ 0.00 & 0.84 $\pm$ 0.02 & 0.68 $\pm$ 0.03 & 0.94 $\pm$ 0.01 \\
soc-Epinions1 & 0.77 $\pm$ 0.03 & 0.96 $\pm$ 0.01 &     --         & 0.64 $\pm$ 0.04 & 0.85 $\pm$ 0.02 \\
soc-Slashdot0902 & 0.82 $\pm$ 0.02 & 0.96 $\pm$ 0.00 &     --         & 0.69 $\pm$ 0.03 & 0.91 $\pm$ 0.01 \\
soc-sign-epinions & 0.76 $\pm$ 0.03 & 0.96 $\pm$ 0.01 &     --         & 0.54 $\pm$ 0.04 & 0.81 $\pm$ 0.02 \\
web-Google & 0.79 $\pm$ 0.02 & 0.91 $\pm$ 0.01 &     --         & 0.10 $\pm$ 0.06 & 0.86 $\pm$ 0.02 \\
web-NotreDame & 0.61 $\pm$ 0.04 & 0.80 $\pm$ 0.02 &     --         & 0.31 $\pm$ 0.06 & 0.44 $\pm$ 0.05 \\
web-Stanford & 0.68 $\pm$ 0.03 & 0.78 $\pm$ 0.02 &     --         & 0.37 $\pm$ 0.05 & 0.65 $\pm$ 0.04 \\
wiki-Vote & 0.93 $\pm$ 0.01 & 0.96 $\pm$ 0.00 & 0.48 $\pm$ 0.05 & 0.71 $\pm$ 0.03 & 0.95 $\pm$ 0.01 \\
\end{tabular} 

\clearpage